\documentclass[a4paper,english]{article}
\usepackage[T1]{fontenc}
\usepackage[latin9]{inputenc}
\usepackage{amsmath}
\usepackage{amssymb}
\usepackage{esint}

\makeatletter

\usepackage{a4wide}
\usepackage{hyperref}

\makeatother

\usepackage{babel}
\begin{document}

%%%%%%%%%%%%%%%%%%%%
\begin{titlepage}
%\begin{flushright} \small TUW--10--20 \end{flushright}
%\markright{\bf TUW--11--??}
%\vspace*{.5cm}
\title{On semiclassical calculation of three-point functions in $AdS_5\times T^{1,1}$}

\author{M.~Michalcik${}^{\star}$, R.~C.~Rashkov${}^{\star,\dagger}$\thanks{e-mail:
rash@hep.itp.tuwien.ac.at.} and M.~Schimpf${}^{\star}$
\ \\ \ \\
${}^{\star}$ Institute for Theoretical Physics, \\ Vienna
University of Technology,\\
Wiedner Hauptstr. 8-10, 1040 Vienna, Austria,
\ \\ \ \\
${}^{\dagger}$  Department of Physics, Sofia
University
\\
5 J. Bourchier Blvd, 1164 Sofia, Bulgaria
}
\date{}
\end{titlepage}

\maketitle
%%%%%%%%%%%%%%%%%%%%%%%%%%%%%%%

\begin{abstract}
Recently there has been progress on the computation of two- and three-point correlation functions with
two ``heavy'' states via semiclassical methods. We extend this analysis to the case of $AdS_5\times
T^{1,1}$, and examine the suggested procedure for the case of several simple string solutions.
By making use of AdS/CFT duality, we derive the relevant correlation functions of operators belonging to
the dual gauge theory.
\end{abstract}

\section{Introduction}

One of the most interesting outputs of theoretical physics of last
years was the conjectured duality between the string theory on $AdS_{5}\times S^{5}$
and $N=4$ Super Yang-Mills field theory \cite{Maldacena1998}. Shortly
after Maldacena's paper there were many checks of this conjecture. 
 Due to the complicated
Green-Schwartz superstring action on curved spacetime,
most of them have been worked out  in the supergravity (SUGRA) approximation. In order to
check the duality, one usually compares various observables e.g. scaling
dimensions, correlation functions. These quantities depend in general
non-trivially on the t'Hooft coupling $\lambda=g_{YM}^{2}N$. A perturbative
description of these observables covers however opposite regions of
the coupling $\lambda$, which makes it difficult to check the duality.
Fortunately there is a small class of operators, the BPS operators,
which have due to the supersymmetry trivial dependence on $\lambda$.
They allow to compare both sides of the duality.  The challenge is
however, to go beyond the BPS case. One way is to consider special
limits of parameters other than $\lambda$ as in the BMN case \cite{Berenstein2002}.
A significant simplification comes when one focuses on a particular
class of observables - the spectrum of energies of single string states.
They correspond to the scaling dimensions of the single trace operators
on the field theory side. The idea of BMN relies on the fact that
for particular string states with large quantum numbers there are
new limits where certain quantum corrections are suppressed. In the
BMN case (for reviews see \cite{Pankiewicz2003,Plefka2004,Sadri2004})
one considers a small closed string moving with large angular momentum
$J\gg1$ around a great circle of $S^{5}$. It represents a particular
semi-classical sector  of near BPS states, which allows in the limit
$J\rightarrow\infty,\:\frac{\lambda}{J^{2}}-\text{fixed}$ for an
exact correspondence between energies of the string states and scaling
dimensions of the SYM operators \cite{Berenstein2002}. This is possible
since one can interpret the states semiclassically as quadratic corrections
to a point-like string running along the great circle (the geodesic)
of the $S^{5}$. In this case the corrections higher than one-loop
are suppressed in the limit $J\rightarrow\infty,\:\frac{\lambda}{J^{2}}-\text{fixed}$
\cite{Frolov2002,Tseytlin2003} . For other, far from BPS, semiclassical
sectors of string states we can still hope  for a similar behavior.
In \cite{Frolov2003} it was noticed that for string states with at
least one large $S^{5}$ spin component $J$ the classical energy
has expansion in powers of $\frac{\lambda}{J^{2}}$ and that the quantum
superstring corrections are suppressed in the limit $J\rightarrow\infty,\:\frac{\lambda}{J^{2}}-\text{fixed}$.
In this limit we can thus use (semi)classical approach knowing that
the energy goes as $E=J+\ldots$. Precise tests of the $AdS/CFT$
correspondence in a non-BPS sector were achieved 
\cite{Beisert2003,Frolov2003b,Arutyunov2003,Beisert2003a,Arutyunov2004}
comparing the expansion of the classical string energy with the perturbative
SYM anomalous dimension.  This is however only half of the story,
since in order to describe a CFT one needs to know the spectrum 
and the three point correlation functions of primary operators. As
we note above, much is known about the spectrum of conformal weights.
 The situation with the three-point functions is however different
and there is still a lot to reveal. Since the conformal invariance
determines the spacetime dependence of the three-point correlation
functions the essential information is stored in their couplings.
Even in the well studied $N=4$ SYM case we still do not know much
about the \emph{generic} three-point couplings. It is due to the
lack of a tool for calculation of the three-point correlation functions 
with massive string states.
The path integral approach to compute the string theory partition
function for heavy string states presented in \cite{Janik:2010gc}
opened a new way to compute the two-point function at strong coupling
for the cases presented in \cite{Tseytlin2003a,Frolov2003,Gubser2002}.
Recently a method to calculate the three-point function at strong
coupling corresponding to two heavy string states and a (light) supergravity
field was presented \cite{Costa:2010rz,Zarembo:2010rr}. 
The method was applied to various cases of known string solutions in $AdS_5\times S^5$
\cite{Zarembo:2010rr}-\cite{Arnaudov:2011wq}, as well as to $AdS_4\times CP^3$ 
\cite{Arnaudov:2010kk} and less supersymmetric
cases \cite{Arnaudov:2011yt}.

In this short note we extend
this successful in $AdS_5\times S^5$ background idea to a less supersymmetric case.
Our background is the $AdS_5$ times the Sasaki-Einstein
space $T^{1,1}$. We calculate the three-point functions of certain gauge theory operators
at strong coupling dual to various simple string solutions.

\section{The two-point functions and the method}

The study of the holography for less supersymmetric theories was initiated in \cite{Klebanov:1998hh} where
a stack of D3 branes was putted on the tip of conifold.
The geometry of such backgrounds was described for instance in \cite{Gauntlett:2004yd,Cvetic:2005ft}.
In this section we briefly describe the method and setup our notations for strings in $AdS_5\times T^{1,1}$ background.
In order to keep track with the already know results we use a more
general metric of the squashed sphere $X_{5}$ rather than the $T^{1,1}$.
This setup allows us to turn to the case of the sphere $S^{5}$ or
the Sasaki-Einstein space $T^{1,1}$ simply by a particular choice
of parameters.  For convenience we use Poincar\'e coordinates to
parametrize the $AdS_{5}$ space, since the boundary is given by $z=0$.
The metric $ds^{2}$ for the target space $AdS_{5}\times X_{5}$ is
thus of the form $ds^{2}=ds_{X}^{2}+ds_{A}^{2}$ with the $X_{5}$
and the $AdS_{5}$ part defined as follows : 
\begin{align}
ds_{X}^{2}= & a\left(d\theta_{1}^{2}+\sin^{2}\left(\theta_{1}\right)d\phi_{1}^{2}
+d\theta_{2}^{2}+\sin^{2}\left(\theta_{2}\right)d\phi_{2}^{2}\right)+b\left(d\psi
+\cos\left(\theta_{1}\right)d\phi_{1}+\cos\left(\theta_{2}\right)d\phi_{2}\right)^{2}\label{eq:metric_X}\\
ds_{A}^{2}= & \frac{dx^{\mu}dx_{\mu}+dz^{2}}{z^{2}}\label{eq:metric_A}
\end{align}
For two special values of the parameters $a,b$ , namely $a=1/4\,,\, b=1/4$
and $a=1/6\,,\, b=1/9$ the metric \eqref{eq:metric_X} corresponds
to $S^{5}$ and $T^{1,1}$ respectively. For both, the target space
and the worldsheet we use Minkowski signature. The action we work
with is the standard Polyakov action 
\begin{equation}
S\left[X,\Phi\right]=-\frac{\sqrt{\lambda}}{4\pi}\int\partial_{a}X^{\mu}
\partial_{b}X^{\nu}G_{\mu\nu}h^{ab}d\sigma d\tau e^{\frac{\phi}{2}}+\ldots=-\frac{\sqrt{\lambda}}{4\pi}\int
\left(\mathcal{L}_{X}+\mathcal{L}_{A}\right)d\sigma d\tau\label{eq:Polyakov_action}
\end{equation}
where $G_{\mu\nu}$ is the target space metric, $\phi$ represents
fluctuations of the dilaton field and dots contain other fields such
as the B-field and fermionic terms. We consider trivial topology
of the worldsheet and use the conformal gauge i.e. the 2-dimensional
metric is just $h^{ab}=diag\left(-1,1\right)$. 

Our approach is based on a recent proposal by Costa et al. \cite{Costa:2010rz}
for the calculation of the three-point function at strong coupling.
We will shortly review their method to set up the notation. Let us
first recall the two-point correlation function of operators dual
to classical string states at strong coupling as described in \cite{Janik:2010gc}
.  

As a warm-up example, let us start with a point-like string, which
we distinguish with the (sub)subscript $P$ when necessary. The action
\eqref{eq:Polyakov_action} for the point-particle case has the form
\begin{equation}
S_{P}\left[X,\mathbf{e},\Phi\right]=\frac{1}{2}\int_{0}^{1}d\tau e^{\frac{\phi}{4}}
\left(\mathbf{e}^{-1}\partial_{\tau}X^{\mu}\partial_{\tau}X^{\nu}G_{\mu\nu}-\mathbf{e}m^{2}\right)\label{eq:point_particle_action}
\end{equation}
where $\phi$ represents fluctuations of the dilaton field and $\mathbf{e}=\mathbf{e}\left(\tau\right)$
is the einbein. For convenience one passes from the einbein formulation
to the modular parameter $s$ (see \cite{Janik:2010gc} for details).
The action \eqref{eq:point_particle_action} on the $AdS$ vacuum
takes the form 
\[
S_{P}\left[X,s\right]\equiv S_{P}\left[X,s,\Phi=0\right]=\frac{1}{2}
\int_{-\frac{s}{2}}^{\frac{s}{2}}d\tau\mathcal{L}_{P}=-\frac{1}{2}\int_{-\frac{s}{2}}^{\frac{s}{2}}
\left(m^{2}-\frac{x'^{2}\left(\tau\right)+z'^{2}\left(\tau\right)}{z^{2}\left(\tau\right)}\right)d\tau
\]
where we assume that the particle is moving along a direction $x$
 and that all distances on the boundary are space-like. A generic
solution to the equations of motion 
\begin{align}
x\left(\tau\right)= & R\tanh\left(\kappa\tau\right)+x_{0}\nonumber \\
z\left(\tau\right)= & \frac{R}{\cosh\left(\kappa\tau\right)}\label{eq:EoM_P}
\end{align}
includes three free parameters $R,\kappa,x_{0}$ , which are further
fixed by the following boundary conditions
\begin{equation}
x\left(-\frac{s}{2}\right)=0\,,\, z\left(-\frac{s}{2}\right)=\epsilon\,,\, 
x\left(\frac{s}{2}\right)=x_{f}\,,\, z\left(\frac{s}{2}\right)=\epsilon.\label{eq:ads_boundary_cond}
\end{equation}
Imposing conditions \eqref{eq:ads_boundary_cond} one finds 
\begin{align*}
z\left(\pm\frac{s}{2}\right)= & \frac{R}{\cosh\left(\kappa\frac{s}{2}\right)}=\epsilon.
\end{align*}
In the limit $\epsilon\ll1$, we approximately get $x_{0}\sim R$
and $R\sim\frac{x_{f}}{2}$ which results in the following relation
for the parameter $\kappa$ 
\begin{equation}
\kappa\approx\frac{2}{s}\log\left(\frac{x_{f}}{\epsilon}\right).\label{eq:kappa_relation}
\end{equation}

Putting all these relations to the action we obtain 
\[
S_{P}\left[\bar{X},s\right]=\frac{1}{2}s\left(\kappa^{2}-m^{2}\right)=\frac{1}{2}s
\left(\frac{4\log^{2}\left(\frac{x_{f}}{\epsilon}\right)}{s^{2}}-m^{2}\right)
\]
where the second equality comes after the substitution of \eqref{eq:kappa_relation}.
With the value of the saddle point%
\footnote{Since we are working in Minkowski signature we get an imaginary saddle
point.%
} 
\begin{equation}
\bar{s}=-\imath\frac{2\log\left(\frac{x_{f}}{\epsilon}\right)}{m}\label{eq:saddle_point}
\end{equation}
 we compute the 2-point function of the following form 
\begin{equation}
\left\langle \mathcal{O}\left(0\right),\mathcal{O}\left(x_{f}\right)\right\rangle 
\approx e^{iS_{P}\left[\bar{X},\bar{s}\right]}=\left(\frac{x_{f}}{\epsilon}\right)^{-2m}.\label{eq:corr_point}
\end{equation}

Before we go on with the 3-point function let us recall that in order
to obtain the correct scaling of the two-point correlation function
for the operator $\mathcal{O}_{A}$ dual to a heavy string field one
has to convolute the generating functional with the wave function
as described in \cite{Janik:2010gc} . Practically it boils down to
a change of the measure in the string path integral, such that we
work with the effective action  
\begin{align}
\tilde{S}\left[X,s\right]=S\left[X,s\right]-\bar{S}\left[X,s\right]= & S\left[X,s\right]
-\int_{-\frac{s}{2}}^{\frac{s}{2}}d\tau\int_{0}^{2\pi}d\sigma\left(\left(\Pi-\Pi_{0}
\right)^{a}\left(\dot{X}-\dot{X}_{0}\right)_{a}+\Pi^{i}\dot{X}_{i}\right)\label{eq:S_tilde_general}\\
\Pi_{0}^{a}= & \frac{1}{2\pi}\int_{0}^{2\pi}d\sigma\Pi^{a}\left(\tau,\sigma\right)
\;,\;\dot{X}_{0}^{a}=\frac{1}{2\pi}\int_{0}^{2\pi}d\sigma\dot{X}^{a}\left(\tau,\sigma\right)\nonumber 
\end{align}
where $\Pi$ are the worldsheet canonical momenta. The sub/super scripts
$\left\{ a,i\right\} $ are indices for the AdS and $X_{5}$ parts
respectively. 

The 3-point correlation function $\left\langle \mathcal{O}_{A}\left(x_{i}\right)
\mathcal{O}_{A}\left(x_{f}\right)\mathcal{D}_{\chi}\left(y\right)\right\rangle $
with $\mathcal{O}_{A}$ being the operator corresponding to a heavy
string field and $\mathcal{D}_{\chi}$ a chiral operator corresponding
to a light supergravity field can be approximated \cite{Costa:2010rz}
as follows 
\begin{equation}
\left\langle \mathcal{O}_{A}\left(x_{i}\right)\mathcal{O}_{A}\left(x_{f}\right)
\mathcal{D}_{\chi}\left(y\right)\right\rangle \approx\frac{I_{\chi}\left[\bar{X},
\bar{s},y\right]}{\left|x_{i}-x_{f}\right|^{2\Delta_{A}}}.\label{eq:3-pt_general}
\end{equation}
Here $\Delta_{A}$ is the dimension of the operator $\mathcal{O}_{A}$
; $x_{i}\,,\, x_{f}$ are the insertion points of the operators $\mathcal{O}_{A}$,
$y$ is the insertion point of the operator $\mathcal{D}_{\chi}$
and $I_{\chi}$ is the interaction term between heavy string fields
and the supergravity field $\chi$ . This interaction term is of the
form 
\[
I_{\chi}\left[X,s,y\right]=\imath\int_{-\frac{s}{2}}^{\frac{s}{2}}d\tau
\int_{0}^{2\pi}d\sigma\left.\frac{\delta S\left[X,s,\Phi\right]}{\delta\chi}
\right|_{\Phi=0}K_{\chi}\left(X\left(\tau,\sigma\right),y\right)
\]

with $K_{\chi}\left(X\left(\tau,\sigma\right),y\right)$ being the
bulk-to-boundary propagator for the field $\chi$. We introduce for
convenience the integrand $i_{\chi}$ such that $I_{\chi}=\int_{\frac{s}{2}}^{\frac{s}{2}}\int_{0}^{2\pi}i_{\chi}d\tau d\sigma$.
In the case of the dilaton field $\left(\chi=\phi\right)$ and the
action \eqref{eq:Polyakov_action} we thus have 
\begin{equation}
i_{\phi}=\frac{\imath}{2}\mathcal{L}K_{\phi}=\frac{3}{2\pi^{2}}\mathcal{L}
\left(\frac{z}{z^{2}+\left(x-y\right)^{2}}\right)^{4},\label{eq:i_phi_general}
\end{equation}
where $\mathcal{L}=\mathcal{L}_{A}+\mathcal{L}_{X}$ is the Lagrangian.
Calculating the interaction term for the point particle and taking
the leading order approximation in the parameter $\epsilon$ we get 

\[
I_{\phi_{P}}\approx-\frac{\imath}{32\pi^{2}\log\left(\frac{x_{f}}{\epsilon}\right)s}
\left(s^{2}m^{2}-4\log^{2}\left(\frac{x_{f}}{\epsilon}\right)\right)\frac{x_{f}^{4}}{\left(x_{f}-y\right)^{4}y^{4}}.
\]
Evaluating the above interaction term at the saddle point \eqref{eq:saddle_point}
we end up with 
\[
I_{\phi_{P}}\approx-\frac{m}{8\pi^{2}}\frac{x_{f}^{4}}{\left(x_{f}-y\right)^{4}y^{4}},
\]
which gives a 3-point function of the form 
\begin{equation}
\left\langle \mathcal{O}_{A}\left(0\right)\mathcal{O}_{A}\left(x_{f}\right)\mathcal{L}
\left(y\right)\right\rangle \approx-\frac{m}{8\pi^{2}}\frac{1}{x_{f}^{2\Delta_{A}-4}y^{4}
\left(x_{f}-y\right)^{4}}.\label{eq:3-pt_point}
\end{equation}

We can check the consistency of our result with the expectation from
the renormalization group as it was proposed in \cite{Costa:2010rz}.
For the coupling $a_{\mathcal{L}AA}$ we obtain 
\[
2\pi^{2}a_{\mathcal{L}AA}=-\lambda\frac{\partial}{\partial\lambda}E\approx-\frac{m}{4}
\]
which is in agreement with the result \eqref{eq:3-pt_point} .

\section{Rotating string models}

In our models we consider a point-like string on the $AdS$ space
i.e. 
\begin{equation}
x=x\left(\tau\right)\,,\, z=z\left(\tau\right)\label{eq:AdS_ansatz}
\end{equation}
and a general spinning string in the $X_{5}$ space \cite{Arutyunov:2003za}
described by 
\begin{gather}
\phi_{1}=\omega_{1}\tau\,,\,\phi_{2}=\omega_{2}\tau\,,\,\theta_{1}=\theta_{1}
\left(\sigma\right)\,,\,\theta_{2}=\theta_{2}\left(\sigma\right)\,,\,\psi=\nu\tau.\label{eq:gansatz}
\end{gather}
Inserting the ansatz \eqref{eq:AdS_ansatz},\eqref{eq:gansatz} into
\eqref{eq:Polyakov_action} we obtain the action $S\left[X,\Phi\right]=-\frac{\sqrt{\lambda}}{4\pi}
\int\left(\mathcal{L}_{X}+\mathcal{L}_{A}\right)d\sigma d\tau$
with 
\begin{equation}
\mathcal{L}_{A}=\frac{\overrightarrow{x}'\left(\tau\right)^{2}+z'\left(\tau\right)^{2}}{z\left(\tau\right)^{2}}\label{eq:AdS_action}
\end{equation}
being the Lagrangian for the $AdS$ part and 
\begin{multline}
\mathcal{L}_{X}=b\left(-\nu^{2}-2\nu\omega_{1}\cos\left(\theta_{1}\left(\sigma\right)
\right)-2\nu\omega_{2}\cos\left(\theta_{2}\left(\sigma\right)\right)\right.\\
\left.-2\omega_{1}\omega_{2}\cos\left(\theta_{1}\left(\sigma\right)\right)\cos\left(\theta_{2}
\left(\sigma\right)\right)\right)-\omega_{1}^{2}\left(b\cos^{2}\left(\theta_{1}\left(\sigma\right)
\right)+a\sin^{2}\left(\theta_{1}\left(\sigma\right)\right)\right)\\
-\omega_{2}^{2}\left(b\cos^{2}\left(\theta_{2}\left(\sigma\right)\right)+a\sin^{2}\left(\theta_{2}
\left(\sigma\right)\right)\right)+a\left(\theta_{1}'\left(\sigma\right)^{2}+\theta_{2}'\left(\sigma\right)^{2}\right).\label{eq:X_action}
\end{multline}
specifying the $X_{5}$ part. Similarly we find the effective action
$\tilde{S}\left[X,s\right]$. According to \eqref{eq:S_tilde_general}
it is enough to write down the {}``subtracted'' action $\bar{S}\left[X,s\right]
=-\frac{\sqrt{\lambda}}{4\pi}\int\left(\bar{\mathcal{L}}_{X}+\bar{\mathcal{L}}_{A}\right)d\sigma d\tau$,
substituting the ansatz \eqref{eq:AdS_ansatz},\eqref{eq:gansatz}
to \eqref{eq:S_tilde_general}. The Lagrangians we obtain are of the
following form 
\begin{align}
\bar{\mathcal{L}}_{A}= & 0\nonumber \\
\bar{\mathcal{L}}_{X}= & -2b\left(\nu^{2}+2\nu\omega_{1}\cos\left(\theta_{1}
\left(\sigma\right)\right)+2\nu\omega_{2}\cos\left(\theta_{2}\left(\sigma\right)\right)\right.\nonumber \\
 & \left.+\omega_{1}^{2}\cos^{2}\left(\theta_{1}\left(\sigma\right)\right)+\omega_{2}^{2}\cos^{2}
\left(\theta_{2}\left(\sigma\right)\right)+2\omega_{1}\omega_{2}\cos\left(\theta_{1}\left(\sigma\right)\right)
\cos\left(\theta_{2}\left(\sigma\right)\right)\right)\label{eq:L_bar_X}\\
 & -2a\left(\omega_{1}^{2}\sin^{2}\left(\theta_{1}\left(\sigma\right)\right)+\omega_{2}^{2}\sin^{2}
\left(\theta_{2}\left(\sigma\right)\right)\right).\nonumber 
\end{align}

There are three conserved charges $J_{\phi_{1}},J_{\phi_{2}},J_{\psi}$
 corresponding to the symmetries of the $X_{5}$ space with their
currents $P_{\phi_{1}},P_{\phi_{2}},P_{\psi}$ satisfying $J_{I}=-\frac{\sqrt{\lambda}}{4\pi}\int_{0}^{2\pi}P_{I}d\sigma$,
 where $I\in\left\{ A,B,R\right\} $. 
Evaluating the currents $P_{I}$ for the Lagrangian $\mathcal{L}=\mathcal{L}_{A}+\mathcal{L}_{X}$
\eqref{eq:AdS_action},\eqref{eq:X_action} we obtain 
\begin{align}
P_{A}=P_{\phi_{1}}= & \frac{\partial\mathcal{L}}{\partial\left(\partial_{\tau}\phi_{1}\right)}
=-2\left(\left(b\cos^{2}\left(\theta_{1}\left(\sigma\right)\right)+a\sin^{2}\left(\theta_{1}
\left(\sigma\right)\right)\right)\phi_{1}'\left(\tau\right)\right)\nonumber \\
 & -2\left(b\cos\left(\theta_{1}\left(\sigma\right)\right)\left(\cos\left(\theta_{2}
\left(\sigma\right)\right)\phi_{2}'\left(\tau\right)+\psi'\left(\tau\right)\right)\right)\nonumber \\
P_{B}=P_{\phi_{2}}= & \frac{\partial\mathcal{L}}{\partial\left(\partial_{\tau}\phi_{2}\right)}
=-2\left(b\cos\left(\theta_{1}\left(\sigma\right)\right)\cos\left(\theta_{2}\left(\sigma\right)
\right)\phi_{1}'\left(\tau\right)\right)\label{eq:charges_generall}\\
 & -2\left(\left(b\cos^{2}\left(\theta_{1}\left(\sigma\right)\right)+a\sin^{2}\left(\theta_{1}
\left(\sigma\right)\right)\right)\phi_{2}'\left(\tau\right)+b\cos\left(\theta_{2}\left(\sigma\right)
\right)\psi'\left(\tau\right)\right)\nonumber \\
P_{R}=P_{\psi}= & \frac{\partial\mathcal{L}}{\partial\left(\partial_{\tau}\psi\right)}
=-2b\left(\cos\left(\theta_{1}\left(\sigma\right)\right)\phi_{1}'\left(\tau\right)
+\cos\left(\theta_{2}\left(\sigma\right)\right)\phi_{2}'\left(\tau\right)+\psi'\left(\tau\right)\right).\nonumber 
\end{align}
 Here and in the following we use \emph{prime} for the derivation
when there is one variable only. The Virasoro constraints can be written
in a concise form 
\begin{align*}
vir_{1}=vir_{1A}+vir_{1X}= & T_{\tau\tau}+T_{\sigma\sigma}\:,\: vir_{2}=vir_{2A}+vir_{2X}=T_{\tau\sigma}
\end{align*}
where $T_{ab}$ is the world-sheet energy-momentum tensor. With the
ansatz \eqref{eq:AdS_ansatz} their explicit form for the $AdS$ part
is 
\begin{align*}
vir_{1A}= & \frac{x'\left(\tau\right)^{2}+z'\left(\tau\right)^{2}}{z\left(\tau\right)^{2}}\:,\: vir_{2A}=0.
\end{align*}
 Similarly using the ansatz \eqref{eq:gansatz} we obtain the $X_{5}$
part as 
\begin{multline*}
vir_{1X}=b\left(\nu^{2}+2\nu\omega_{1}\cos\left(\theta_{1}\left(\sigma\right)\right)+\omega_{1}^{2}\cos^{2}
\left(\theta_{1}\left(\sigma\right)\right)+2\nu\omega_{2}\cos\left(\theta_{2}\left(\sigma\right)\right)\right.\\
\left.+2\omega_{1}\omega_{2}\cos\left(\theta_{1}\left(\sigma\right)\right)\cos\left(\theta_{2}\left(\sigma\right)
\right)+\omega_{2}^{2}\cos^{2}\left(\theta_{2}\left(\sigma\right)\right)\right)\\
+a\left(\omega_{1}^{2}\sin^{2}\left(\theta_{1}\left(\sigma\right)\right)+\omega_{2}^{2}\sin^{2}\left(\theta_{2}
\left(\sigma\right)\right)+\theta_{1}'\left(\sigma\right)^{2}+\theta_{2}'\left(\sigma\right)^{2}\right)
\end{multline*}
and 
\[
vir_{2X}=0.
\]

In the following part we analyze particular string solutions.

\subsection{single spin case }

The single spin rotating string extended in the $\theta_{1}$ direction
and rotating in the $\phi_{1}$ is given by the following consistent
truncation of \eqref{eq:gansatz} 
\begin{equation}
\theta_{1}\left(\sigma\right)=\theta\left(\sigma\right)\,,\,\theta_{2}\left(\sigma\right)=0\,,\,\omega_{1}=\omega\,
,\,\omega_{2}=0\,,\,\nu=0.\label{eq:single_spin_ansatz}
\end{equation}
Substituting the ansatz \eqref{eq:single_spin_ansatz} to \eqref{eq:AdS_action},\eqref{eq:X_action},\eqref{eq:L_bar_X}
we find the following Lagrangians 
\begin{align}
\mathcal{L}_{1}= & -\frac{x'\left(\tau\right)^{2}+z'\left(\tau\right)^{2}}{z\left(\tau\right)^{2}}
-\omega^{2}\left(b\cos^{2}\left(\theta\left(\sigma\right)\right)+a\sin^{2}\left(\theta\left(\sigma\right)
\right)\right)+a\theta'\left(\sigma\right)^{2}\label{eq:lagrangian_single_spin}\\
\bar{\mathcal{L}}_{1}= & -2\omega^{2}\left(b\cos^{2}\left(\theta\left(\sigma\right)\right)+a\sin^{2}
\left(\theta\left(\sigma\right)\right)\right).\nonumber 
\end{align}
They result in the action 
\begin{align}
\tilde{S}_{1}\left[\bar{X},s\right] & =-\frac{\sqrt{\lambda}}{4\pi}\left(-2\pi\kappa^{2}s+s\int_{0}^{2\pi}
\left(\omega^{2}\left(b\cos^{2}\left(\theta\left(\sigma\right)\right)+a\sin^{2}\left(\theta\left(\sigma\right)
\right)\right)+a\theta'\left(\sigma\right)^{2}\right)d\sigma\right)\label{eq:S_cir1}
\end{align}
where we already put the solution for the AdS part \eqref{eq:EoM_P}
. The subscript 1 indicates the single spin case. Rewriting the action
\eqref{eq:S_cir1} in a convenient form and substituting $\kappa$
from \eqref{eq:kappa_relation} we obtain 
\begin{align}
\tilde{S}_{1}\left[\bar{X},s\right] & =-\frac{\sqrt{\lambda}}{4\pi}\left(-2\pi\kappa^{2}s+sI_{C}\right)
=-\frac{\sqrt{\lambda}}{4\pi}\left(-\frac{8\pi\log^{2}\left(\frac{x_{f}}{\epsilon}\right)}{s}+sI_{C}\right)\label{eq:S_tilde_single}
\end{align}
where $I_{C}$ is given by 
\begin{equation}
I_{C}=\int_{0}^{2\pi}\left(\omega^{2}\left(b\cos^{2}\left(\theta\left(\sigma\right)\right)+a\sin^{2}
\left(\theta\left(\sigma\right)\right)\right)+a\theta'\left(\sigma\right)^{2}\right)d\sigma.\label{eq:IC_cir1}
\end{equation}
The conserved charges $J_{I}$ for the single spin Lagrangian \eqref{eq:lagrangian_single_spin}
 takes the following form: 
\begin{align}
J_{A}= & \frac{\sqrt{\lambda}}{2\pi}\omega\int_{0}^{2\pi}d\sigma\left(b\cos^{2}\left(\theta\left(\sigma\right)
\right)+a\sin^{2}\left(\theta\left(\sigma\right)\right)\right)\nonumber \\
J_{B}= & \frac{\sqrt{\lambda}}{2\pi}\omega b\int_{0}^{2\pi}d\sigma\cos\left(\theta\left(\sigma\right)\right)
\label{eq:charges_fold}\\
J_{R}= & \frac{\sqrt{\lambda}}{2\pi}\omega b\int_{0}^{2\pi}d\sigma\cos\left(\theta\left(\sigma\right)\right).\nonumber 
\end{align}

Next we proceed with calculation of the 2-point and the 3-point function
keeping the $\theta\left(\sigma\right)$ unsolved for the moment.
We calculate the 2-point function using the saddle point approximation.
Since the integral \eqref{eq:IC_cir1} is independent of the modular
parameter $s$, the saddle point $\bar{s}$ of the action \eqref{eq:S_tilde_single}
is of the form 
\begin{equation}
\bar{s}=-\frac{2\imath\sqrt{2\pi}\log\left(\frac{x_{f}}{\epsilon}\right)}{\sqrt{I_{C}}}.\label{eq:fold_saddle_point}
\end{equation}
The action $\tilde{S}_{1}$ at the above saddle point becomes 
\[
\tilde{S}_{1}\left[\bar{X},\bar{s}\right]=\imath\sqrt{\frac{2}{\pi}}\sqrt{\lambda}\sqrt{I_{C}}\log\left(\frac{x_{f}}{\epsilon}\right),
\]

and results in the 2-point function of the form 
\begin{equation}
\left\langle \mathcal{O}\left(0\right),\mathcal{O}\left(x_{f}\right)\right\rangle \approx e^{i\tilde{S}_{1}
\left[\bar{X},\bar{s}\right]}=\left(\frac{x_{f}}{\epsilon}\right)^{-\sqrt{\frac{2}{\pi}}\sqrt{\lambda}\sqrt{I_{C}}}.\label{eq:2_pt_fold}
\end{equation}

In order to obtain the three-point function one has to calculate the
interaction term $I_{\phi}$ as indicated in \eqref{eq:3-pt_general}
(see \cite{Costa:2010rz} for details). Inserting the Lagrangian \eqref{eq:lagrangian_single_spin}
to the relation \eqref{eq:i_phi_general}, the integrand $i_{\phi}$
for the single spin string becomes 
\begin{multline*}
i_{\phi_{1}}=\frac{3\imath x_{f}^{4}\left(\frac{x_{f}}{\epsilon}\right)^{\frac{8\tau}{s}}}{2\pi^{2}s^{2}
\left(y^{2}+\left(x_{f}-y\right)^{2}\left(\frac{x_{f}}{\epsilon}\right)^{\frac{4\tau}{s}}\right)^{4}}\left(s^{2}
\omega^{2}\left(-a-b+\left(a-b\right)\cos\left(2\theta\left(\sigma\right)\right)\right)\right.\\
\left.-8\log^{2}\left(\frac{x_{f}}{\epsilon}\right)+2as^{2}\theta'^{2}\left(\sigma\right)\right).
\end{multline*}
Using a small $\epsilon$ approximation for the interaction term $I_{\phi}$
we find 
\begin{multline*}
I_{\phi_{1}}\approx\left(\int_{0}^{2\pi}\left(s^{2}\omega^{2}\left(a+b-\left(a-b\right)\cos\left(2\theta
\left(\sigma\right)\right)\right)+8\log^{2}\left(\frac{x_{f}}{\epsilon}\right)-2as^{2}\theta'^{2}\left(\sigma\right)\right)d\sigma\right)\\
\frac{\imath x_{f}^{4}\sqrt{\lambda}}{64\pi^{3}s\left(x_{f}-y\right)^{4}y^{4}\log\left(\frac{x_{f}}{\epsilon}\right)},
\end{multline*}
which at the saddle point \eqref{eq:fold_saddle_point} turns to 
\begin{equation}
I_{\phi_{1}}\approx\frac{\imath x_{f}^{4}\sqrt{\lambda}}{32\pi^{3}\left(x_{f}-y\right)^{4}y^{4}\kappa}
\int_{0}^{2\pi}\left(2\kappa^{2}+\omega^{2}\left(a+b-\left(a-b\right)\cos\left(2\theta\left(\sigma\right)\right)
\right)-2a\theta'^{2}\left(\sigma\right)\right)d\sigma.\label{eq:I_phi_cir_temp}
\end{equation}
Using the Virasoro constraint \eqref{eq:vir_cir1} and the $J_{A}$
charge \eqref{eq:charges_fold} the interaction term \eqref{eq:I_phi_cir_temp}
becomes 

\[
I_{\phi_{1}}\approx\frac{\imath\left(\kappa^{2}\sqrt{\lambda}+\omega J_{A}\right)}{4\pi^{2}\kappa}
\frac{x_{f}^{4}}{\left(x_{f}-y\right)^{4}y^{4}}.
\]
According to \eqref{eq:3-pt_general} we find the following three-point
function 
\begin{equation}
\left\langle \mathcal{O}_{A}\left(0\right)\mathcal{O}_{A}\left(x_{f}\right)\mathcal{L}\left(y\right)\right\rangle 
\approx\frac{\imath\left(\kappa^{2}\sqrt{\lambda}+\omega J_{A}\right)}{4\pi^{2}\kappa}
\frac{1}{x_{f}^{2\Delta_{A}-4}y^{4}\left(x_{f}-y\right)^{4}}.\label{eq:3_pt_cir1}
\end{equation}

The two \eqref{eq:2_pt_fold} and three \eqref{eq:3_pt_cir1} point
function we obtained above include the yet unsolved $\theta\left(\sigma\right)$
in a more or less hidden form. In order to solve for $\theta\left(\sigma\right)$
we use the first Virasoro constraint 
\begin{equation}
\kappa^{2}+b\omega^{2}+\left(a-b\right)\omega^{2}\sin^{2}\left(\theta\left(\sigma\right)\right)+a\theta'^{2}\left(\sigma\right)=0,\label{eq:vir_cir1}
\end{equation}
where all parameters except of $\kappa$ are real, so we demand $\kappa^{2}+b\omega^{2}<0$.
Introducing the following parameters 
\begin{equation}
c=\imath\frac{\sqrt{\kappa^{2}+b\omega^{2}}}{\sqrt{a}}\:,\: 
k=-\frac{\left(a-b\right)\omega^{2}}{\kappa^{2}+b\omega^{2}}\label{eq:c-k_substitution}
\end{equation}
we get a convenient form of the equation \eqref{eq:vir_cir1} 
\begin{align}
\frac{\partial\theta\left(\sigma\right)}{\partial\sigma}= & c\sqrt{1-k\sin^{2}
\left(\theta\left(\sigma\right)\right)}.\label{eq:virasoro_generic_diffEq}
\end{align}
It is necessary at this point to distinguish between two cases depending
on  $k$ : 
\begin{enumerate}
\item $k>1$ where $\theta\left(\sigma\right)$ reaches a turning point
and thus describes a folded string.  
\item $k<1$ where the function $\theta\left(\sigma\right)$ is monotonous
and thus describes a circular string. Such a string winds around the
great circle of the $S^{2}$.
\end{enumerate}
We analyze these cases separately below. For completeness we note
that both cases are constrained by positive value of the parameter
$k$.

\paragraph{Folded string}

We start our analysis with the folded string case. In order to bring
the equation \eqref{eq:virasoro_generic_diffEq} with $k>1$ into
a convenient form we use the transformation 
\begin{align}
k\sin^{2}\theta\left(\sigma\right)= & \sin^{2}X\left(\sigma\right),\label{eq:fold_subst}
\end{align}
leading to the following equation 
\[
\frac{\partial X\left(\sigma\right)}{\partial\sigma}=c\sqrt{k}\sqrt{1-\frac{1}{k}\sin^{2}\left(X\left(\sigma\right)\right)}.
\]
The solution to this differential equation subject to the condition
$X\left(0\right)=0$, is expressed in terms of the Jacobi Amplitude
($am\left(z,m\right)=\mathbb{F}^{-1}\left(z,m\right)$) as 
\begin{equation}
X\left(\sigma\right)=am\left(c\sqrt{k}\sigma,\frac{1}{k}\right).\label{eq:X_fold_sol}
\end{equation}
The closed string periodicity condition 
\begin{equation}
X\left(0\right)=X\left(2\pi\right)+2\pi\label{eq:X_periodicity_condition}
\end{equation}
demands the constant $c$ to be  
\begin{align}
c= & -\frac{2\mathbb{K}\left(\frac{1}{k}\right)}{\sqrt{k}\pi},\label{eq:fold_periodicity_cond}
\end{align}
bringing up the final form of $X\left(\sigma\right)$ 
\begin{equation}
X\left(\sigma\right)=am\left(-\frac{2\mathbb{K}\left(\frac{1}{k}\right)}{\pi}\sigma,\frac{1}{k}\right).\label{eq:X_fold_sol1}
\end{equation}
 The expression \eqref{eq:fold_periodicity_cond} with the definition
of parameters \eqref{eq:c-k_substitution} results in a relation between
$\kappa$ and $\omega$ of the form 
\begin{align}
\mathbb{K}\left(\frac{1}{k}\right)=\mathbb{K}\left(-\frac{\kappa^{2}+b\omega^{2}}{\left(a-b\right)
\omega^{2}}\right)= & \frac{\pi}{2}\omega\frac{\sqrt{a-b}}{\sqrt{a}}.\label{eq:fold_K_rule}
\end{align}
 Inserting the solution \eqref{eq:X_fold_sol1} and the substitution
\eqref{eq:fold_subst} into the single-spin relations for conserved
charges \eqref{eq:charges_fold} we obtain 
\begin{align}
J_{B} & =J_{R}=\frac{\sqrt{\lambda}b\pi\omega}{2\mathbb{K}\left(\frac{1}{k}\right)}=\frac{\sqrt{a}
b\sqrt{\lambda}}{\sqrt{a-b}}\nonumber \\
J_{A} & =\frac{\sqrt{\lambda}\omega\left(a\mathbb{K}\left(\frac{1}{k}\right)-\left(a-b\right)
\mathbb{E}\left(\frac{1}{k}\right)\right)}{\mathbb{K}\left(\frac{1}{k}\right)}=a\sqrt{\lambda}
\omega-\frac{2}{\pi}\sqrt{\lambda}\sqrt{a}\sqrt{a-b}\mathbb{E}\left(\frac{1}{k}\right)\label{eq:JA_fold}
\end{align}
where in the last equality we already used the relation \eqref{eq:fold_K_rule}.
Expressing $\kappa$ from \eqref{eq:fold_K_rule} we calculate the
energy 
\begin{align}
E=\frac{\sqrt{\lambda}}{2\pi}\int_{0}^{2\pi}-\imath\kappa d\sigma= & \sqrt{\lambda}\omega
\sqrt{b+\left(a-b\right)\mathbb{K}^{-1}\left(\frac{\pi\omega\sqrt{a-b}}{2\sqrt{a}}\right)}.\label{eq:energy_fold}
\end{align}
We note that in order to get the dispersion relation from \eqref{eq:energy_fold}
one needs to substitute $\omega$ from the $J_{A}$ charge \eqref{eq:JA_fold}.
It is however a transcendental equation because $k$ depends on $\omega$
and $\kappa$ \eqref{eq:c-k_substitution}, hence there is no algebraic
expression for the dispersion relation.

Let us return to the two \eqref{eq:2_pt_fold} and the three \eqref{eq:3_pt_cir1}
-point function calculated for the single spin string. In order to
complete the two-point function calculation we need to express the
square root $\sqrt{I_{C}}$ of the integral \eqref{eq:IC_cir1} in
terms of constants and conserved charges. It is  the same as having
the dispersion relation, which however can not be solved algebraically
in this case. A similar argument applies for the three-point function
\eqref{eq:3_pt_cir1}, where in order to express it with charges and
constants only, one needs to find the relation for $\omega/\kappa$.
It again involves transcendental equation and therefore can not be
obtained explicitly.

In order to compare our result with the expectation from the renormalization
flow equation \cite{Costa:2010rz} we need to calculate the derivative
$\frac{\partial}{\partial\lambda}E$. Since the energy \eqref{eq:energy_fold}
is given implicitly we proceed as follows : 
\begin{align}
2\pi^{2}a_{\mathcal{L}AA}= & -\lambda\frac{\partial}{\partial\lambda}E=\lambda\frac{\partial}{\partial\lambda}
\left(\sqrt{\lambda}\imath\kappa\right)=\imath\frac{\sqrt{\lambda}}{2}\left(\kappa+2\lambda\frac{\partial\kappa}{\partial\lambda}
\right)\label{eq:renorm_flow_ch_fold}
\end{align}
where $E$ is the energy and $a_{\mathcal{L}AA}$ the coupling. Using
\eqref{eq:c-k_substitution} we express the derivative $\frac{\partial\kappa}{\partial\lambda}$
in terms of the other parameters 
\begin{equation}
\frac{\partial\kappa}{\partial\lambda}=\frac{\kappa^{2}+b\omega^{2}}{2\left(a-b\right)
\kappa\omega^{2}}\frac{\partial k}{\partial\lambda}+\frac{\kappa}{\omega}\frac{\partial\omega}{\partial\lambda},\label{eq:dkappa_fold}
\end{equation}
and from the relation \eqref{eq:fold_K_rule} we obtain $\frac{\partial k}{\partial\lambda}$
of the form 
\begin{equation}
\frac{\partial k}{\partial\lambda}=-\frac{\sqrt{a-b}\left(k-1\right)k\pi
\frac{\partial\omega}{\partial\lambda}}{\sqrt{a}k\mathbb{E}\left(\frac{1}{k}\right)-
\left(k-1\right)\mathbb{K}\left(\frac{1}{k}\right)}.\label{eq:dk_fold}
\end{equation}
 Having \eqref{eq:dk_fold} the last derivative we need to express
is $\frac{\partial\omega}{\partial\lambda}$. Differentiating the
$J_{A}$ charge \eqref{eq:JA_fold} we find 
\begin{equation}
\frac{\partial\omega}{\partial\lambda}=\frac{\sqrt{a}\left(\sqrt{a}\pi\omega-2\sqrt{a-b}\mathbb{E}
\left(\frac{1}{k}\right)\right)\left(k\mathbb{E}\left(\frac{1}{k}\right)-\left(k-1\right)\mathbb{K}
\left(\frac{1}{k}\right)\right)}{2\pi\lambda\left(-\left(a+b\left(k-1\right)\right)\mathbb{E}
\left(\frac{1}{k}\right)+b\left(k-1\right)\mathbb{K}\left(\frac{1}{k}\right)\right)}.\label{eq:domega_fold}
\end{equation}
 Finally substituting \eqref{eq:dk_fold} and \eqref{eq:domega_fold}
into \eqref{eq:dkappa_fold} and using the relations \eqref{eq:JA_fold},\eqref{eq:fold_K_rule}
we find the same result as from the saddle point approximation, i.e.
\[
a_{\mathcal{L}AA}=\imath\frac{\kappa^{2}\sqrt{\lambda}+\omega J_{A}}{4\pi^{2}\kappa}.
\]

\paragraph{circular string case}

As indicated in the analysis of the Virasoro constraint \eqref{eq:virasoro_generic_diffEq}
the circular string case is realized when $k<1$ . For this range
of the parameter $k$ we get a solution to $\theta$ in the form 
\begin{equation}
\theta\left(\sigma\right)=am\left(c\sigma,k\right).\label{eq:cir1_sol}
\end{equation}
 Due to the periodicity condition, which in this case includes winding
around the great circle of $S^{2}$ we have 
\begin{align}
am\left(2\pi c,k\right)= & am\left(0,k\right)+2\pi,\label{eq:cir1_periodicity_cond}
\end{align}
which brings up the following relation between $\kappa$ and $\omega$
\begin{align}
\mathbb{K}\left(k\right)=\mathbb{K}\left(-\frac{\left(a-b\right)\omega^{2}}{\kappa^{2}
+b\omega^{2}}\right)=\frac{\pi}{2}c= & \imath\frac{\pi}{2}\frac{\sqrt{\kappa^{2}+b\omega^{2}}}{\sqrt{a}}.\label{eq:cir1_K_rule}
\end{align}
The solution for $\theta\left(\sigma\right)$ is then 
\begin{equation}
\theta\left(\sigma\right)=am\left(\frac{2\mathbb{K}\left(k\right)}{\pi}\sigma,k\right).\label{eq:theta_cir1_sol1}
\end{equation}

In order to find an explicit solution to the conserved charges \eqref{eq:charges_fold}
we insert the solution \eqref{eq:theta_cir1_sol1} to the relations
for charges \eqref{eq:charges_fold}, and using the relation \eqref{eq:cir1_K_rule}
we obtain 
\begin{align}
J_{B} & =J_{R}=0\nonumber \\
J_{A} & =-\frac{\kappa^{2}\sqrt{\lambda}}{\omega}-\frac{2\imath\sqrt{a}\sqrt{\lambda}\sqrt{\kappa^{2}
+b\omega^{2}}\mathbb{E}\left(k\right)}{\pi\omega}.\label{eq:cir1_A_charge}
\end{align}
An explicit relation for energy in the circular string case is more
complicated, because the relation \eqref{eq:cir1_K_rule} includes
$\kappa$ on the right hand side. It is therefore not possible to
express $\kappa$ from \eqref{eq:cir1_K_rule} in an algebraic way,
therefore the energy 
\begin{align}
E= & \frac{\sqrt{\lambda}}{2\pi}\int_{0}^{2\pi}-\imath\kappa d\sigma=-\sqrt{\lambda}\imath\kappa,\label{eq:cir1_energy}
\end{align}
is subjected to the relation for $\kappa$ \eqref{eq:cir1_K_rule}. 

Similar to the folded case having an explicit relation for the 2-point
function means having the dispersion relation. In order to find it
one has to solve a set of equations \{\eqref{eq:cir1_energy},\eqref{eq:cir1_K_rule},\eqref{eq:cir1_A_charge}\},
which again can not be done algebraically. The same argumentation
applies to the three-point function \eqref{eq:3_pt_cir1} where one
has to find the ratio $\omega/\kappa$ , which can not be done explicitly.
The difference in the 2-point and the 3-point function between the
folded and the circular single spin string is in the relation between
$\kappa$ and $\omega$ \eqref{eq:cir1_K_rule} and the $J_{A}$ charge
\eqref{eq:cir1_A_charge}.

We proceed comparing the 3-point function to the expectation from
the renormalization flow, starting with the expression 
\begin{equation}
2\pi^{2}a_{\mathcal{L}AA}=-\lambda\frac{\partial}{\partial\lambda}E=\lambda\frac{\partial}{\partial\lambda}
\left(\sqrt{\lambda}\imath\kappa\right)=\imath\frac{\sqrt{\lambda}}{2}\left(\kappa+2\lambda
\frac{\partial\kappa}{\partial\lambda}\right).\label{eq:RF_cir}
\end{equation}
In order to compute $\frac{\partial\kappa}{\partial\lambda}$ we differentiate
the parameter $k$ \eqref{eq:c-k_substitution} and the relation \eqref{eq:cir1_K_rule}
obtaining the following equations 
\begin{align*}
\frac{\partial k}{\partial\lambda}\left(\kappa^{2}+b\omega^{2}\right)^{2}= & 2\left(a-b\right)\kappa\omega m
\left(\omega\frac{\partial\kappa}{\partial\lambda}-\kappa\frac{\partial\omega}{\partial\lambda}\right)\\
\frac{\frac{\partial k}{\partial\lambda}\left(\mathbb{E}\left(k\right)-\left(1-k\right)\mathbb{K}\left(k\right)
\right)}{2\left(1-k\right)k}= & \frac{\imath\pi\left(\kappa\frac{\partial\kappa}{\partial\lambda}
+b\omega\frac{\partial\omega}{\partial\lambda}\right)}{2\sqrt{a}\sqrt{\kappa^{2}+b\omega^{2}}}.
\end{align*}
Solving them for $\frac{\partial\kappa}{\partial\lambda}\,\text{and}\,\frac{\partial k}{\partial\lambda}$
we find the relations 
\begin{align}
\frac{\partial\kappa}{\partial\lambda}= & \left(\frac{\kappa}{\omega}-\frac{\imath\pi\left(\kappa^{2}+a\omega^{2}
\right)\sqrt{\kappa^{2}+b\omega^{2}}}{2\kappa\omega\sqrt{a}\mathbb{E}\left(k\right)}\right)
\frac{\partial\omega}{\partial\lambda}\label{eq:dkappa_cir}\\
\frac{\partial k}{\partial\lambda}= & -\frac{\imath\pi\omega\left(a-b\right)\left(\kappa^{2}
+a\omega^{2}\right)}{\sqrt{a}\left(\kappa^{2}+b\omega^{2}\right)^{3/2}\mathbb{E}\left(k\right)}
\frac{\partial\omega}{\partial\lambda},\label{eq:dk_cir}
\end{align}
which substituted to $\frac{\partial J_{A}}{\partial\lambda}=0$ results
in  
\begin{equation}
\frac{\partial\omega}{\partial\lambda}=\frac{\omega\mathbb{E}\left(k\right)-\sqrt{a}\pi\kappa^{2}
-2\imath a\sqrt{k^{2}+b\omega^{2}}\mathbb{E}\left(k\right)}{\pi\lambda\left(2\sqrt{a}\kappa^{2}\mathbb{E}\left(k\right)
-\imath\pi\left(\kappa^{2}+a\omega^{2}\right)\sqrt{k^{2}+b\omega^{2}}\right)}.\label{eq:domega_cir}
\end{equation}
 Finally substituting \eqref{eq:domega_cir} and \eqref{eq:dkappa_cir}
to \eqref{eq:RF_cir} and using the relation \eqref{eq:cir1_K_rule}
we end up with 
\begin{align*}
a_{\mathcal{L}AA} & =\frac{\imath\left(\kappa\sqrt{\lambda}+\frac{\omega}{\kappa}J_{A}\right)}{4\pi^{2}}
=\frac{-E+\imath\frac{\omega}{\kappa}J_{A}}{4\pi^{2}}
\end{align*}
which coincides with the result \eqref{eq:3_pt_cir1}.

\subsection{Two spin folded string}

In the following case we analyze a string extended in the $\theta_{1}$
and rotating in the $\Phi_{1}$ and $\Psi$ direction, described by
a consistent truncation of \eqref{eq:gansatz} of the form 
\[
\theta_{1}\left(\sigma\right)=\theta\left(\sigma\right)\,,\,\theta_{2}\left(\sigma\right)=0\,,
\,\omega_{1}=\omega\,,\,\omega_{2}=0\,,\,\nu\neq0\implies\Psi=\nu\tau.
\]
For the above ansatz the Lagrangians $\mathcal{L}$ \eqref{eq:AdS_action},\eqref{eq:X_action}
and $\bar{\mathcal{L}}$ \eqref{eq:L_bar_X} becomes 
\begin{align}
\mathcal{L}_{2}= & -\frac{x'\left(\tau\right)^{2}+z'\left(\tau\right)^{2}}{z\left(\tau\right)^{2}}
-\omega^{2}\left(b\cos^{2}\left(\theta\left(\sigma\right)\right)+a\sin^{2}\left(\theta\left(\sigma\right)
\right)\right)+a\theta'\left(\sigma\right)^{2}-b\nu^{2}-2b\nu\omega\cos\left(\theta\left(\sigma\right)\right)\label{eq:cir2_L}\\
\bar{\mathcal{L}}_{2}= & -2\left[\omega^{2}\left(b\cos^{2}\left(\theta\left(\sigma\right)\right)+a\sin^{2}
\left(\theta\left(\sigma\right)\right)\right)+b\nu^{2}+2b\nu\omega\cos\left(\theta\left(\sigma\right)\right)\right].\label{eq:cir2_L_bar}
\end{align}
 We use the subscript 2 to distinguish the two spin case when necessary.
The action $\tilde{S}_{2}$ with the solution for the $AdS$ part
\eqref{eq:EoM_P} already inserted, takes the form 
\begin{multline}
\tilde{S}_{2}\left[\bar{X},s\right]=-\frac{\sqrt{\lambda}}{4\pi}\left(-2\pi s\left(\kappa^{2}-b\nu^{2}
\right)+s\int_{0}^{2\pi}\left(\omega^{2}\left(b\cos^{2}\left(\theta\left(\sigma\right)\right)+a\sin^{2}
\left(\theta\left(\sigma\right)\right)\right)\right.\right.\\
\left.\left.+2b\nu\omega\cos\left(\theta\left(\sigma\right)\right)+a\theta'\left(\sigma\right)^{2}\right)
d\sigma\right).\label{eq:S_cir2}
\end{multline}
In order to find the solution for $\theta\left(\sigma\right)$ we
use the first Virasoro constraint obtaining
\begin{align}
-a\theta'^{2}\left(\sigma\right)= & \kappa^{2}+b\nu^{2}+a\omega^{2}+2b\nu\omega
\cos\left(\theta\left(\sigma\right)\right)-\left(a-b\right)\omega^{2}\cos^{2}\left(\theta\left(\sigma\right)\right).\label{eq:theta_cir2}
\end{align}
The above equation includes a term linear in $\cos\left(\theta\left(\sigma\right)\right)$,
therefore it is useful to rewrite \eqref{eq:theta_cir2} in a more
convenient form 
\begin{equation}
\theta'^{2}\left(\sigma\right)=\frac{a-b}{a}\omega^{2}\left(\cos\left(\theta\left(\sigma\right)\right)
-\alpha\right)\left(\cos\left(\theta\left(\sigma\right)\right)-\beta\right),\label{eq:theta_cir2_root}
\end{equation}
where $\alpha$ and $\beta$ are roots of the right hand side of \eqref{eq:theta_cir2_root}
regarded as a polynomial in $\cos\left(\theta\right)$. A folded string
we analyze, extends between a minimal and a maximal angle $\theta\in\left\langle \theta_{min},
\theta_{max}\right\rangle \subset\left(-\frac{\pi}{2},\frac{\pi}{2}\right)$,
which in terms of $\cos\left(\theta\right)$ means $\cos\left(\theta\right)\in\left(0,1\right\rangle $.
First we note that without loss of generality we set $\theta\left(0\right)=0$
and that a folded string requires at least one of the roots $\alpha,\beta$
to be the turning point. Regarding the right hand side of \eqref{eq:theta_cir2_root}
as a quadratic polynomial in $\cos\left(\theta\right)$, we find that
in order to reach a turning point i.e. to have a \emph{non-negative
}region we must set $\alpha,\beta\in\left(-1,1\right)$. Due to $\nu\neq0$
the roots $\alpha$ and $\beta$ are different, so we choose $\beta>\alpha$
. In order to have a turning point we must set $0<\beta<1$ and because
of the consistency of the folded string we assume also $-1<\alpha<0$.
With all these constraints we find that the turning point is at $\cos\left(\theta_{turn}\right)=\beta$
and that the folded string extends between $\pm\theta_{turn}$ i.e.
$\theta\in\left(-\theta_{turn},\theta_{turn}\right)$.  The roots
$\alpha$ and $\beta$ have the following form 
\begin{align}
\alpha= & \frac{b\nu-\sqrt{\left(a-b\right)\kappa^{2}+ab\nu^{2}+a\left(a-b\right)\omega^{2}}}{\left(a-b\right)\omega}\nonumber \\
\beta= & \frac{b\nu+\sqrt{\left(a-b\right)\kappa^{2}+ab\nu^{2}+a\left(a-b\right)\omega^{2}}}{\left(a-b\right)\omega}.\label{eq:cir2_roots}
\end{align}

Solving the equation \eqref{eq:theta_cir2_root} for $\theta\left(\sigma\right)$
we get the following expression 
\begin{align}
\sqrt{\frac{a-b}{a}}\omega\sigma & =\int_{0}^{\theta}\frac{\mathbf{d}\theta}
{\sqrt{\left(\cos\left(\theta\left(\sigma\right)\right)-\alpha\right)\left(\cos\left(\theta
\left(\sigma\right)\right)-\beta\right)}}=I_{1}\left(\theta\right)-I_{1}\left(0\right)\label{eq:theta_cir2_sol}
\end{align}
giving the solution to $\theta\left(\sigma\right)$ only implicitly.
This is because the integral $I_{1}\left(\theta\right)$ \eqref{eq:integrals_I1_indef}
is up to a multiplicative factor the elliptic integral of the first
kind. The folded string periodicity condition gives the following
relation 
\begin{align}
2\pi\sqrt{\frac{a-b}{a}}\omega & =4\int_{0}^{\theta_{turn}}\frac{\mathbf{d}\theta}{\sqrt{\left(\cos\left(\theta\left(\sigma\right)\right)
-\alpha\right)\left(\cos\left(\theta\left(\sigma\right)\right)-\beta\right)}}=4\left(I_{1}\left(acrcos\left(\beta\right)\right)-I_{1}
\left(0\right)\right).\label{eq:cir2_periodicity_cond_calc}
\end{align}
Substituting for $I_{1}\left(\theta\right)$ from \eqref{eq:integrals_I1_indef}
the relation \eqref{eq:cir2_periodicity_cond_calc} becomes 
\begin{equation}
\pi\omega=4\sqrt{\frac{a}{a-b}}\frac{\mathbb{K}\left(k\right)}{\sqrt{\left(1-\alpha\right)\left(1+\beta\right)}},\label{eq:cir2_periodicity_condition}
\end{equation}
where $k$ is defined as 
\begin{equation}
k=\frac{\left(1+\alpha\right)\left(1-\beta\right)}{\left(1-\alpha\right)\left(1+\beta\right)}.\label{eq:cir2_k_rule}
\end{equation}

Inserting the two spin Lagrangian \eqref{eq:cir2_L} in the relations
for currents \eqref{eq:charges_generall}, one finds the following
expressions for conserved charges  
\begin{align}
J_{A}= & \frac{\sqrt{\lambda}}{2\pi}\int_{0}^{2\pi}d\sigma\left[\omega\left(b\cos^{2}
\left(\theta\left(\sigma\right)\right)+a\sin^{2}\left(\theta\left(\sigma\right)\right)\right)
+b\nu\cos\left(\theta\left(\sigma\right)\right)\right]\nonumber \\
J_{R}=J_{B}= & \frac{\sqrt{\lambda}}{2\pi}b\int_{0}^{2\pi}d\sigma\left(\nu
+\omega\cos\left(\theta\left(\sigma\right)\right)\right).\label{eq:cir2_charges_def}
\end{align}
The solution to $\theta\left(\sigma\right)$ \eqref{eq:theta_cir2_sol}
is defined implicitly, so we can not insert it directly to the relations
\eqref{eq:cir2_charges_def}. It is therefore better to make a substitution
$\sigma\rightarrow\theta\left(\sigma\right)$ in the above integrals
based on \eqref{eq:theta_cir2_root}, which for an arbitrary function
$f\left(\sigma\right)$ becomes 
\[
\int_{0}^{\frac{\pi}{2}}f\left(\sigma\right)d\sigma\rightarrow\int_{0}^{\theta_{turn}}f
\left(\theta\left(\sigma\right)\right)\frac{d\theta}{\theta'}=\int_{0}^{arccos\left(\beta\right)}
\frac{f\left(\theta\left(\sigma\right)\right)d\theta}{\sqrt{\left(\cos\left(\theta\left(\sigma\right)\right)
-\alpha\right)\left(\cos\left(\theta\left(\sigma\right)\right)-\beta\right)}}.
\]
There are three integrals entering the calculation, which we denote
as $I_{1},I_{2},I_{3}$ with the following definition: 
\begin{align}
I_{1}= & \int_{0}^{arccos\left(\beta\right)}\frac{1}{\sqrt{\left(\cos\left(\theta\right)-\alpha\right)
\left(\cos\left(\theta\right)-\beta\right)}}d\theta\nonumber \\
I_{2}= & \int_{0}^{arccos\left(\beta\right)}\frac{\cos\left(\theta\right)}{\sqrt{\left(\cos\left(\theta\right)
-\alpha\right)\left(\cos\left(\theta\right)-\beta\right)}}d\theta\label{eq:Integrals}\\
I_{3}= & \int_{0}^{arccos\left(\beta\right)}\sqrt{\left(\cos\left(\theta\right)-\alpha\right)
\left(\cos\left(\theta\right)-\beta\right)}d\theta.\nonumber 
\end{align}
The reader can find their expression in the appendix \eqref{eq:integrals_explicit}.
Rewriting relations for the charges \eqref{eq:cir2_charges_def} using
the integrals \eqref{eq:Integrals} we obtain 
\begin{align}
J_{A} & =-\frac{\sqrt{\lambda}}{\pi\omega^{2}}\sqrt{\frac{a}{a-b}}\left(\left(\kappa^{2}+b\nu^{2}
\right)I_{1}+\omega\left(\left(a-b\right)\omega I_{3}+b\nu I_{2}\right)\right)\nonumber \\
J_{B} & =\frac{\sqrt{\lambda}}{\pi\omega}2b\sqrt{\frac{a}{a-b}}\left(\nu I_{1}+\omega I_{2}\right).\label{eq:cir2_charges_int}
\end{align}
Substituting the integrals $I_{i}$ from \eqref{eq:integrals_explicit},
the charges \eqref{eq:cir2_charges_int} becomes 
\begin{align}
J_{A}= & -\frac{\sqrt{\lambda}}{\pi\omega^{2}}\sqrt{\frac{a}{a-b}}\frac{2}{\sqrt{\left(1-\alpha\right)
\left(1+\beta\right)}}\left(\left(a-b\right)\left(1-\alpha\right)\left(1+\beta\right)\omega^{2}\mathbb{E}\left(k\right)\right.\nonumber \\
 & +2\left(\kappa^{2}+b\nu^{2}-b\nu\omega+\left(a-b\right)\alpha\left(1+\beta\right)\omega^{2}\right)\mathbb{K}\left(k\right)\nonumber \\
 & \left.+2\omega\left(2b\nu-\left(a-b\right)\left(\alpha+\beta\right)\omega\right)\mathbb{PI}\left(n,k\right)\right)\nonumber \\
J_{B}= & \frac{\sqrt{\lambda}}{\pi\omega}2b\sqrt{\frac{a}{a-b}}\frac{2}{\sqrt{\left(1-\alpha\right)\left(1+\beta\right)}}
\left(\left(\nu-\omega\right)\mathbb{K}\left(k\right)+2\omega\mathbb{PI}\left(n,k\right)\right).\label{eq:cir2_charges_sol}
\end{align}

In order to calculate the energy 
\begin{equation}
E=\frac{\sqrt{\lambda}}{2\pi}\int_{0}^{2\pi}-\imath\kappa d\sigma=-\sqrt{\lambda}\imath\kappa,\label{eq:cir2_energy}
\end{equation}
one needs to express $\kappa$ from the relation \eqref{eq:cir2_periodicity_condition}.
It is again a complicated transcendental equation, which therefore
determines $\kappa$ only implicitly. The dispersion relation given
via the system of equations \{\eqref{eq:cir2_energy},\eqref{eq:cir2_periodicity_condition},\eqref{eq:cir2_charges_sol}\}
can not be solved in an algebraic way either. 

The calculation of the 2-point function follows the same steps as
in cases discussed before. Using the integral 

\[
I_{C2}=\int_{0}^{2\pi}\left(\omega^{2}\left(b\cos^{2}\left(\theta\left(\sigma\right)\right)+a\sin^{2}
\left(\theta\left(\sigma\right)\right)\right)+2b\omega\nu\cos\left(\theta\left(\sigma\right)\right)+a\theta'\left(\sigma\right)^{2}\right)d\sigma,
\]
the solution \eqref{eq:EoM_P} and the relation for $\kappa$ \eqref{eq:kappa_relation}
we rewrite the action \eqref{eq:S_cir2} in the following concise
form 
\begin{align}
\tilde{S}_{2}\left[\bar{X},s\right] & =-\frac{\sqrt{\lambda}}{4\pi}\left(-\frac{8\pi\log^{2}
\left(\frac{x_{f}}{\epsilon}\right)}{s}+s\left(2\pi b\nu^{2}+I_{C2}\right)\right).\label{eq:cir2_S2}
\end{align}
The above action has the following saddle point 
\begin{equation}
\bar{s}=-\frac{2\imath\sqrt{2\pi}\log\left(\frac{x_{f}}{\epsilon}\right)}{\sqrt{2\pi b\nu^{2}+I_{C2}}}.\label{eq:cir2_saddle_point}
\end{equation}
Evaluating the action \eqref{eq:cir2_S2} at the saddle point \eqref{eq:cir2_saddle_point}
we find 
\[
\tilde{S}_{2}\left[\bar{X},\bar{s}\right]=\imath\sqrt{\frac{2}{\pi}}\sqrt{\lambda}\sqrt{2\pi b\nu^{2}+I_{C2}}\log\left(\frac{x_{f}}{\epsilon}\right),
\]

which results in the following 2-point function 
\[
\left\langle \mathcal{O}\left(0\right),\mathcal{O}\left(x_{f}\right)\right\rangle \approx e^{i\tilde{S}_{2}
\left[\bar{X},\bar{s}\right]}=\left(\frac{x_{f}}{\epsilon}\right)^{-\sqrt{\frac{2}{\pi}}\sqrt{\lambda}\sqrt{2\pi b\nu^{2}+I_{C2}}}.
\]

The integrand $i_{\phi}$ \eqref{eq:i_phi_general} of the interaction
term  becomes in the two spin case

\begin{multline*}
i_{\phi_{2}}=-\frac{3\imath x_{f}^{4}\left(\frac{x_{f}}{\epsilon}\right)^{\frac{8\tau}{s}}}{2\pi^{2}s^{2}\left(y^{2}
+\left(x_{f}-y\right)^{2}\left(\frac{x_{f}}{\epsilon}\right)^{\frac{4\tau}{s}}\right)^{4}}\left(s^{2}\omega^{2}
\left(a+b-\left(a-b\right)\cos\left(2\theta\left(\sigma\right)\right)\right)\right.\\
\left.+s^{2}\left(2b\nu^{2}+4b\nu\omega\cos\left(\theta\left(\sigma\right)\right)-2a\theta'^{2}\left(\sigma\right)
\right)+8\log^{2}\left(\frac{x_{f}}{\epsilon}\right)\right).
\end{multline*}
A small $\epsilon$ approximation of the interaction term results
in the following expression 
\begin{multline*}
I_{\phi_{2}}\approx\left(\int_{0}^{2\pi}\left(s^{2}\omega^{2}\left(a-\left(a-b\right)\cos^{2}\left(\theta
\left(\sigma\right)\right)\right)+4\log^{2}\left(\frac{x_{f}}{\epsilon}\right)\right.\right.\\
\left.\left.+s^{2}\left(b\nu^{2}+2b\nu\omega\cos\left(\theta\left(\sigma\right)\right)-a\theta'^{2}
\left(\sigma\right)\right)\right)\mathbb{d}\sigma\right)\frac{\imath x_{f}^{4}\sqrt{\lambda}}{32\pi^{3}s
\left(x_{f}-y\right)^{4}y^{4}\log\left(\frac{x_{f}}{\epsilon}\right)}.
\end{multline*}
Inserting the saddle point \eqref{eq:cir2_saddle_point} and using
\eqref{eq:theta_cir2} we obtain 
\begin{equation}
I_{\phi_{2}}\approx\frac{\imath x_{f}^{4}\sqrt{\lambda}}{8\pi^{3}\left(x_{f}-y\right)^{4}y^{4}\kappa}
\int_{0}^{2\pi}\left(\kappa^{2}+b\nu^{2}+2b\nu\omega\cos\left(\theta\left(\sigma\right)\right)+\omega^{2}
\left(a-\left(a-b\right)\cos^{2}\left(\theta\left(\sigma\right)\right)\right)\right)d\sigma.\label{eq:cir2_I_Phi_temp}
\end{equation}
In order to rewrite the interaction term \eqref{eq:cir2_I_Phi_temp}
in terms of conserved charges we use the relations \eqref{eq:cir2_charges_def}
and obtain 
\[
I_{\phi_{2}}\approx\frac{\imath\left(\kappa^{2}\sqrt{\lambda}+\omega J_{A}+\nu J_{B}\right)}{4\pi^{2}\kappa}
\frac{x_{f}^{4}}{\left(x_{f}-y\right)^{4}y^{4}},
\]
which results in the three-point function of the form 
\begin{equation}
\left\langle \mathcal{O}_{A}\left(0\right)\mathcal{O}_{A}\left(x_{f}\right)\mathcal{L}\left(y\right)
\right\rangle \approx\frac{\imath\left(\kappa^{2}\sqrt{\lambda}+\omega J_{A}+\nu J_{B}\right)}{4\pi^{2}
\kappa}\frac{1}{x_{f}^{2\Delta_{A}-4}y^{4}\left(x_{f}-y\right)^{4}}.\label{eq:cir2_3-pt_result}
\end{equation}

In the last part of this section we compare our result of the three-point
function \eqref{eq:cir2_3-pt_result} to the expectation from the renormalization
flow equation. Since energy is given by an implicit relation we start
with the following expression
\begin{equation}
2\pi^{2}a_{\mathcal{L}AA}=-\lambda\frac{\partial}{\partial\lambda}E=\lambda\frac{\partial}{\partial\lambda}
\left(\sqrt{\lambda}\imath\kappa\right)=\imath\frac{\sqrt{\lambda}}{2}\left(\kappa+2\lambda
\frac{\partial\kappa}{\partial\lambda}\right).\label{eq:cir2_RF}
\end{equation}
In order to find $\frac{\partial\kappa}{\partial\lambda}$ we use
the following equations 
\begin{align}
\frac{\partial J_{A}}{\partial\lambda}= & 0\:,\:\frac{\partial J_{B}}{\partial\lambda}=0\label{eq:cir2_RF_J}\\
\pi\frac{\partial\omega}{\partial\lambda}= & \frac{\partial}{\partial\lambda}\left(4\sqrt{\frac{a}{a-b}}
\frac{\mathbb{K}\left(k\right)}{\sqrt{\left(1-\alpha\right)\left(1+\beta\right)}}\right),\label{eq:cir2_RF_K}
\end{align}
where the last one comes from the differentiation of \eqref{eq:cir2_periodicity_condition}.
It is very useful to simplify the above set of equations \{\eqref{eq:cir2_RF_J},\eqref{eq:cir2_RF_K}\}
before the actual calculation. We do this expressing the three elliptic
functions $\mathbb{K}\left(k\right),\mathbb{E}\left(k\right),\mathbb{PI}\left(n,k\right)$
from the relations \eqref{eq:cir2_charges_sol},\eqref{eq:cir2_periodicity_condition}
as 
\begin{align}
\mathbb{K}\left(k\right)= & \frac{1}{4}\pi\omega\sqrt{\frac{a-b}{a}}\sqrt{\left(1-\alpha\right)\left(1+\beta\right)}\nonumber \\
\mathbb{E}\left(k\right)= & \frac{\pi}{2\sqrt{a\left(a-b\right)\left(1-\alpha\right)\left(1+\beta\right)}}
\left(\frac{-J_{A}}{\sqrt{\lambda}}+\frac{1}{2}\omega\left(b\left(\alpha-\beta\right)+a\left(2-\alpha+\beta\right)\right)\right)\nonumber \\
\mathbb{PI}\left(n,k\right)= & \frac{\pi\sqrt{a\left(a-b\right)\left(1-\alpha\right)\left(a+\beta\right)}}{8ab}
\left(\frac{J_{B}}{\sqrt{\lambda}}-b\left(\nu-\omega\right)\right),\label{eq:cir2_subs_rules}
\end{align}
and substituting them back to \{\eqref{eq:cir2_RF_J},\eqref{eq:cir2_RF_K}\}.
The equations \eqref{eq:cir2_RF_J} becomes 
\begin{multline}
b\left(1-\beta\right)\lambda\left(2J_{A}-\left(b\left(\alpha-\beta\right)+a\left(2-\alpha+\beta\right)\right)
\sqrt{\lambda}\omega\right)\frac{\partial\alpha}{\partial\lambda}\\
-\left(1-\alpha\right)\left(b\lambda\left(2J_{A}+\left(b\left(\alpha-\beta\right)+a\left(2-\alpha+\beta\right)
\right)\sqrt{\lambda}\omega\right)\frac{\partial\beta}{\partial\lambda}\right)\\
-\left(a-b\right)\left(\alpha-\beta\right)\left(1-\beta\right)\left(J_{B}+2b\lambda^{\frac{3}{2}}
\left(\frac{\partial\nu}{\partial\lambda}-\frac{\partial\omega}{\partial\lambda}\right)\right)=0\label{eq:cir2_RF_JA}
\end{multline}

and 
\begin{multline}
\left(1+\alpha\right)\lambda\left(2J_{A}-\left(-b\left(\alpha-\beta\right)+a\left(2+\alpha-\beta\right)
\right)\sqrt{\lambda}\omega\right)\frac{\partial\beta}{\partial\lambda}\\
-\left(1-\beta\right)\lambda\left(2J_{A}+\left(b\left(\alpha-\beta\right)-a\left(2+\alpha-\beta\right)
\right)\sqrt{\lambda}\omega\right)\frac{\partial\alpha}{\partial\lambda}\\
-2\left(1+\alpha\right)\left(1-\beta\right)\left(J_{A}+\left(b\left(\alpha-\beta\right)+a\left(2-\alpha+\beta\right)\right)
\lambda^{\frac{3}{2}}\right)\frac{\partial\omega}{\partial\lambda}=0\label{eq:cir2_RF_JB}
\end{multline}
and the third one \eqref{eq:cir2_RF_K} turns to 
\begin{multline}
\left(-1+\beta^{2}\right)\left(2J_{A}+\left(-b\alpha\left(\alpha-\beta\right)-a\left(2-\left(\alpha
-\beta\right)\alpha\right)\right)\sqrt{\lambda}\omega\right)\frac{\partial\alpha}{\partial\lambda}\\
-\left(-1+\alpha^{2}\right)\left(2J_{A}-\left(-b\beta\left(\alpha-\beta\right)+a\left(2+\left(\alpha
-\beta\right)\beta\right)\right)\sqrt{\lambda}\omega\right)\frac{\partial\beta}{\partial\lambda}\\
+\left(-1+\alpha^{2}\right)\left(-1+\beta^{2}\right)2\left(a-b\right)\left(\alpha-\beta\right)
\sqrt{\lambda}\frac{\partial\omega}{\partial\lambda}=0.\label{eq:cir2_RD_K}
\end{multline}
We find $\frac{\partial\alpha}{\partial\lambda}\text{ and }\frac{\partial\beta}{\partial\lambda}$
differentiating the relation \eqref{eq:theta_cir2_root} as 
\begin{align}
\frac{\partial\alpha}{\partial\lambda}= & \frac{1}{\left(a-b\right)\left(\alpha-\beta\right)\omega^{2}}
\left(2\kappa\frac{\partial\kappa}{\partial\lambda}+\left(b\left(\alpha-\beta\right)+a\left(\alpha+\beta
\right)\right)\omega\frac{\partial\nu}{\partial\lambda}\right.\nonumber \\
 & \left.+\left(b\alpha\left(\alpha-\beta\right)+a\left(2-\left(\alpha-\beta\right)\alpha\right)\right)\omega
\frac{\partial\omega}{\partial\lambda}\right)\nonumber \\
\frac{\partial\beta}{\partial\lambda}= & -\frac{1}{\left(a-b\right)\left(\alpha-\beta\right)\omega^{2}}
\left(2\kappa\frac{\partial\kappa}{\partial\lambda}+\left(-b\left(\alpha-\beta\right)+a\left(\alpha
+\beta\right)\right)\omega\frac{\partial\nu}{\partial\lambda}\right.\nonumber \\
 & \left.+\left(-b\beta\left(\alpha-\beta\right)+a\left(2+\left(\alpha-\beta\right)\beta\right)\right)
\omega\frac{\partial\omega}{\partial\lambda}\right).\label{eq:cir2_RF_alpha_beta}
\end{align}
Substituting $\frac{\partial\alpha}{\partial\lambda}\text{ and }\frac{\partial\beta}{\partial\lambda}$
from \eqref{eq:cir2_RF_alpha_beta} and the roots $\alpha\text{ and }\beta$
from \eqref{eq:cir2_roots} into the set of equations \{\eqref{eq:cir2_RF_JA},\eqref{eq:cir2_RF_JB},\eqref{eq:cir2_RD_K}\}
we obtain a form, which includes the three terms $\frac{\partial\kappa}{\partial\lambda}\,,
\,\frac{\partial\nu}{\partial\lambda}\,,\,\frac{\partial\omega}{\partial\lambda}$
, the conserved charges and  parameters $\kappa,\nu,\omega$. Solving
this complicated set of equations for $\frac{\partial\kappa}{\partial\lambda}$
we find 
\begin{equation}
W\left(-J_{B}\nu-J_{A}\omega+2\kappa\lambda^{\frac{3}{2}}\frac{\partial\kappa}{\partial\lambda}\right)=0,\label{eq:cir2_kappa_eq}
\end{equation}
where the term $W$ is of the following form 
\begin{multline*}
W=-2a^{2}\omega^{3}+\kappa^{2}\sqrt{\left(a-b\right)\kappa^{2}+ab\nu^{2}+a\left(a-b\right)\omega^{2}}\\
+2a\omega\left(-\kappa^{2}+b\left(\omega^{2}-\nu^{2}\right)+\omega\sqrt{\left(a-b\right)\kappa^{2}+ab\nu^{2}+a\left(a-b\right)\omega^{2}}\right)\\
+b\left(2\kappa^{2}\omega+\left(\nu-\omega\right)\left(\nu+\omega\right)\sqrt{\left(a-b\right)\kappa^{2}+ab\nu^{2}+a\left(a-b\right)\omega^{2}}\right).
\end{multline*}
Assuming $W\neq0$ we find the following expression for $\frac{\partial\kappa}{\partial\lambda}$
\begin{equation}
\frac{\partial\kappa}{\partial\lambda}=\frac{J_{B}\nu+J_{A}\omega}{2\kappa\lambda^{\frac{3}{2}}}.\label{eq:cir2_RF_kappa}
\end{equation}
Finally substituting $\frac{\partial\kappa}{\partial\lambda}$ from
\eqref{eq:cir2_RF_kappa} to \eqref{eq:cir2_RF} we obtain 
\[
a_{\mathcal{L}AA}=\frac{\imath\left(\kappa^{2}\sqrt{\lambda}+J_{B}\nu+J_{A}\omega\right)}{4\pi^{2}\kappa}
\]
which coincides with the solution for the three-point function \eqref{eq:cir2_3-pt_result}
.

\section{Conclusions}

 The holographic conjecture has been tested in many cases
 and impressive results about anomalous dimensions of the gauge theory operators, 
integrable structures, etc., and crucial properties of various gauge theories at
 strong coupling have been established. One of the main challenges ahead is to find efficient
 methods for calculation of the correlation functions.

 While discovering a semiclassical trajectory controlling the leading contribution
 to the three-point correlator of arbitrary ``heavy`` vertex operators is so far an unsolved problem,
 we have seen that one can use the trajectory for the correlation function of two ``heavy`` operators, which is straightforward to find,
to compute the correlator containing two ''heavy'' and one ''light'' states at strong coupling 
\cite{Zarembo:2010rr, Costa:2010rz}.
 The approach based on insertion of vertex operators was put forward in Ref. \cite{Roiban:2010,BT:2010}
 where the authors also suggested that the same method applies to higher n-point
 correlation functions with two ``heavy`` and n-2 ``light'' operators. Namely, the
 semiclassical expression for n-point correlator should be given by a product of
 ``light`` vertex operators calculated on the worldsheet surface determined by the
 ''heavy'' operator insertions.
 In the present paper we considered string theory on $AdS_5 \times T^{1,1}$ and computed
 three-point correlation functions of two ``heavy`` (string) and one ''light`` 
(super-gravity) states at strong coupling, applying the ideas of Ref. \cite{Roiban:2010} 
for calculation of correlation functions using vertex operators for the corresponding states. 
We examined the method in the case of a folded string solution with one and two spins in
$T^{1,1}$ part of the geometry. The solutions we use are the most simple ones and 
correspond to particular gauge theory operators discussed in the text. 
We also checked the consistency of the obtained structure constants of the correlators with
those obtained as $2\pi^2 a_{\mathcal L AA}=\lambda\partial E/\partial\lambda$ as suggested in \cite{Roiban:2010} and
obtained complete agreement. This was possible because the dispersion relations for the
string solutions are simple enough. 
Keeping parameters $a$ and $b$ of the $T^{1,1}$  metric free gives the option
to compare the results from our considerations, namely strings in $AdS_5\times T^{1,1}$, with
the case when the geometry is $AdS_5\times S^5$. 

The correlations functions for the case of giant magnons and single spikes is much more complicated.
This can be seen also just looking at the dispersion relations for these solutions 
\cite{Benvenuti:2008bd}, which give a transcendental equation for the energy. Certainly, 
along with the generalization of the method to include more heavy operators, the study of the correlation
functions in less supersymmetric gauge theories at strong coupling remains a big challenge.

\appendix
\numberwithin{equation}{section}

\section{The integrals}

For completeness we first write the indefinite integrals as 
\begin{align}
I_{1}\left(\theta\right)= & \int\frac{1}{\sqrt{\left(\cos\left(\theta\right)-\alpha\right)
\left(\cos\left(\theta\right)-\beta\right)}}d\theta\nonumber \\
I_{2}\left(\theta\right)= & \int\frac{\cos\left(\theta\right)}{\sqrt{\left(\cos\left(\theta\right)
-\alpha\right)\left(\cos\left(\theta\right)-\beta\right)}}d\theta\label{eq:indefinite_int}\\
I_{3}\left(\theta\right)= & \int\sqrt{\left(\cos\left(\theta\right)-\alpha\right)\left(\cos\left(\theta\right)-\beta\right)}d\theta.\nonumber 
\end{align}
It is then clear that the integrals $I_{i}$ defined in \eqref{eq:Integrals}
are just 
\begin{equation}
I_{i}=\left.I_{i}\left(\theta\right)\right|_{0}^{arccos\left(\beta\right)}.\label{eq:def_integral}
\end{equation}
The explicit form of the integrals \eqref{eq:indefinite_int} is the
following 
\begin{align}
I_{1}\left(\theta\right)= & \frac{-2\mathbb{F}\left(arcsin\left(\frac{\left(1-\alpha\right)
\left(\beta-\cos\theta\right)}{\left(\beta-\alpha\right)\left(1-\cos\theta\right)}\right),
\frac{2\left(\beta-\alpha\right)}{\left(1-\alpha\right)\left(1+\beta\right)}\right)}{\sqrt{\left(\alpha-1\right)
\left(1+\beta\right)}}\label{eq:integrals_I1_indef}\\
I_{2}\left(\theta\right)= & \frac{-2}{\sqrt{\left(1+\beta\right)\left(1-\alpha\right)}}\left[\mathbb{F}
\left(arcsin\left(\sqrt{\frac{1+\beta}{1-\beta}}\tan\left(\frac{\theta}{2}\right)\right),\frac{\left(1-\beta\right)
\left(1+\alpha\right)}{\left(1+\beta\right)\left(1-\alpha\right)}\right)\right.\nonumber \\
 & \left.+2\mathbb{PI}\left(\frac{\beta-1}{\beta+1},-arcsin\left(\sqrt{\frac{1+\beta}{1-\beta}}\tan
\left(\frac{\theta}{2}\right)\right),\frac{\left(1-\beta\right)\left(1+\alpha\right)}{\left(1+\beta\right)
\left(1-\alpha\right)}\right)\right]\nonumber \\
I_{3}\left(\theta\right)= & \frac{1}{\sqrt{\left(1-\alpha\right)\left(1+\beta\right)}}\left[\left(1-\alpha\right)
\left(1+\beta\right)\mathbb{E}\left(arcsin\left(\sqrt{\frac{1+\beta}{1-\beta}}\tan\left(\frac{\theta}{2}\right)\right)
,\frac{\left(1-\beta\right)\left(1+\alpha\right)}{\left(1+\beta\right)\left(1-\alpha\right)}\right)\right.\nonumber \\
 & +2\alpha\left(1+\beta\right)\mathbb{F}\left(arcsin\left(\sqrt{\frac{1+\beta}{1-\beta}}\tan\left(\frac{\theta}{2}
\right)\right),\frac{\left(1-\beta\right)\left(1+\alpha\right)}{\left(1+\beta\right)\left(1-\alpha\right)}\right)\nonumber \\
 & +2\left(\alpha+\beta\right)\mathbb{PI}\left(\frac{\beta-1}{\beta+1},-arcsin\left(\sqrt{\frac{1+\beta}{1-\beta}}
\tan\left(\frac{\theta}{2}\right)\right),\frac{\left(1-\beta\right)\left(1+\alpha\right)}{\left(1+\beta\right)
\left(1-\alpha\right)}\right)\nonumber \\
 & \left.+\left(1-\alpha\right)\sqrt{\left(1+\beta\right)\left(\alpha-\cos\theta\right)\left(\beta-\cos\theta\right)}
\tan\left(\frac{\theta}{2}\right)\right].\nonumber 
\end{align}
The integrals $I_{1},I_{2},I_{3}$ \eqref{eq:def_integral} thus becomes
\begin{align}
I_{1}= & \frac{2\mathbb{K}\left(\frac{\left(1-\beta\right)\left(1+\alpha\right)}{\left(1+\beta\right)
\left(1-\alpha\right)}\right)}{\sqrt{\left(1-\alpha\right)\left(1+\beta\right)}}\nonumber \\
I_{2}= & -2\frac{\mathbb{K}\left(\frac{\left(1-\beta\right)\left(1+\alpha\right)}{\left(1+\beta\right)
\left(1-\alpha\right)}\right)-2\mathbb{PI}\left(\frac{\beta-1}{\beta+1},\frac{\left(1-\beta\right)
\left(1+\alpha\right)}{\left(1+\beta\right)\left(1-\alpha\right)}\right)}{\sqrt{\left(1-\alpha\right)
\left(1+\beta\right)}}\label{eq:integrals_explicit}\\
I_{3}= & \frac{1}{\sqrt{\left(1-\alpha\right)\left(1+\beta\right)}}\left[\left(1-\alpha\right)
\left(1+\beta\right)\mathbb{E}\left(\frac{\left(1-\beta\right)\left(1+\alpha\right)}{\left(1+\beta\right)
\left(1-\alpha\right)}\right)\right.\nonumber \\
 & \left.+2\alpha\left(1+\beta\right)\mathbb{K}\left(\frac{\left(1-\beta\right)\left(1+\alpha\right)}{\left(1+\beta\right)
\left(1-\alpha\right)}\right)+2\left(\alpha+\beta\right)\mathbb{PI}\left(\frac{\beta-1}{\beta+1},\frac{\left(1-\beta\right)
\left(1+\alpha\right)}{\left(1+\beta\right)\left(1-\alpha\right)}\right)\right]\nonumber 
\end{align}

\providecommand{\href}[2]{#2}\begingroup\raggedright
\endgroup

%\bibliographystyle{utphys}
%\bibliography{papers}

\end{document}